\documentclass[conference]{IEEEtran}
\usepackage[backend=biber,style=ieee, maxnames=3]{biblatex}
\usepackage{flushend}
\usepackage{hyperref}
\AtBeginBibliography{\small}
\input epsf
\usepackage{graphicx}
\usepackage{xcolor}

\begin{document}

\title{\LARGE Paving the Way towards 800 Gbps Quantum-Secured\\Optical Channel Deployment in Mission-Critical Environments}
\author{\authorblockN{Marco Pistoia\authorrefmark{1}, Omar Amer\authorrefmark{1}, Monik R.~Behera\authorrefmark{1}, Joseph A. Dolphin\authorrefmark{2}, James F. Dynes\authorrefmark{2}, Benny John\authorrefmark{1}, \\ Paul A. Haigh\authorrefmark{2}, Yasushi Kawakura\authorrefmark{3}, David H.~Kramer\authorrefmark{1}, Jeffrey Lyon\authorrefmark{1}, Navid Moazzami\authorrefmark{4}, \\ Tulasi D. Movva\authorrefmark{1}, Antigoni Polychroniadou\authorrefmark{1}, Suresh Shetty\authorrefmark{1}, Greg Sysak\authorrefmark{4}, Farzam Toudeh-Fallah\authorrefmark{1}, \\ Sudhir Upadhyay\authorrefmark{1}, Robert I. Woodward\authorrefmark{2}, Andrew J. Shields\authorrefmark{2} \\  \\
{\authorrefmark{1}JPMorgan Chase \& Co.}~~~~~~~~~~{\authorrefmark{2}Toshiba Europe Ltd}~~~~~~~~~~{\authorrefmark{3}Toshiba America, Inc}~~~~~~~~~~{\authorrefmark{4}Ciena, Inc}}
}

\DeclareRobustCommand*{\IEEEauthorrefmark}[1]{%
  \raisebox{0pt}[0pt][0pt]{\textsuperscript{\scriptsize\ensuremath{#1}}}}

\maketitle

\begin{abstract}
This article describes experimental research studies conducted towards understanding the implementation aspects of high-capacity quantum-secured optical channels in mission-critical metro-scale operational environments using Quantum Key Distribution (QKD) technology. To the best of our knowledge, this is the first time that an 800 Gbps quantum-secured optical channel --- along with several other Dense Wavelength Division Multiplexed (DWDM) channels on the C-band and multiplexed with the QKD channel on the O-band ---  was established at distances up to 100 km, with secret key-rates relevant for practical industry use cases. In addition, during the course of these trials, transporting a blockchain application over this established channel was utilized as a demonstration of securing a financial transaction in transit over a quantum-secured optical channel. The findings of this research pave the way towards the deployment of QKD-secured optical channels in high-capacity, metro-scale, mission-critical operational environments, such as Inter-Data Center Interconnects.
\end{abstract}

\section{Introduction}
Quantum Key Distribution (QKD) is a well-known symmetric-key distribution method that does not rely on computational assumptions.  Rather, it leverages principles of quantum mechanics to provide communication channels with unconditional security, mathematically proven by multiple studies \cite{Shor-Preskill,Lo-Chau,Koashi}. This article mainly concerns itself with the implementation, deployment and integration of QKD technologies into real-world network infrastructures. For a detailed discussion of the theory behind QKD, as well as its security, challenges, and a survey of state-of-the-art real-world tests, see \cite{alleaume2014using,gyongyosi2019survey, AmerSurvey, OpticalSurvey}. That said, we briefly cover here some of the previous efforts in validating the real-world applicability of QKD technologies. 

Although experimental trials have previously been conducted to investigate QKD technology, they have been done on a limited scale or under conditions that do not represent real-world, high-capacity operational environments. In 2005 researchers at DARPA deployed the first real world QKD network in which they achieved 1 kbps Secret-Key Rate (SKR) at 10 km distance over a non-multiplexed dedicated quantum channel \cite{elliott2005current}. In \cite{PeevM2009TSqk}, a SKR of 3.1 kbps  with Quantum Bit Error Rate (QBER) of 2.6\%  over 33 km of non-multiplexed dedicated quantum channel was achieved as part of the SECOQC project which realized a multi-platform QKD network operating in Vienna. Further, researchers in \cite{sasaki2011field} achieved 18 kbps at 90 km on a non-multiplexed dedicated quantum channel on a link in the Tokyo QKD network in 2011. As will be discussed later, the ability to multiplex both DWDM optical channels and quantum channel are of the utmost importance in real-world operational environments. Therefore, these works cannot be considered for such deployments. 

Indeed there have also been previous efforts to multiplex the optical data channels with the quantum channel. It was first shown that it was possible to do so in \cite{townsend1997simultaneous}, where a quantum channel was multiplexed with a 1.2 Gbps classical channel over 28 km. In \cite{Eraerds2010QuantumKD} researchers followed up on this work by multiplexing 4 DWDM channels with a quantum channel, in which those 4 channels were used for key distillation and 1 Gbps encrypted communication. This work resulted in a SKR of 11 bps at a maximum distance of 50 km. In \cite{kumar2015coexistence} a CV-QKD channel multiplexed with one DWDM channel was demonstrated. In this demonstration the single DWDM channel was run at -3 dBm and at 75 km a key rate of 0.49 kbps was achieved. In \cite{Dynes1} two 100 Gbps DWDM channels (at 1529.55 and 1529.94 nm) were multiplexed with the QKD quantum channel on 1547.72 nm. There a simulated study showed that a quantum channel on the C-band multiplexed with multiple DWDM channels totaling a bandwidth of about 2.5 Tbps could result in secret key-rates on the order of 1 kbps at a distance of 75km. In \cite{mao2018integrating} multiple 200 Gbps and 100 Gbps channels were multiplexed with the quantum channel running at 1310 nm. For this experiment, the polarization encoding Decoy-State BB84 protocol QKD was utilized. During this trial, a 5.5 kbps SKR at 66 km of G652 fiber was achieved.

This article describes in detail a joint experimental research study conducted by JPMorgan Chase, Toshiba and Ciena at JPMorgan Chase's Optical Transport Lab in Columbus, Ohio aimed at understanding the implementation and deployment aspects of an 800 Gbps QKD-secured optical channel in mission-critical metro-scale operational environments. The test bed for this research study was carefully designed to mimic such environments.

To the best of our knowledge, this research study is the first one to demonstrate an 800 Gbps QKD-secured optical channel along with several other Dense Wavelength Division Multiplexed (DWDM) channels on the C-band, multiplexed with the QKD quantum channel on the O-band, at distances up to 100 km. At a 70 km distance over a Corning SMF-28 optical fiber, this system achieved a SKR of 66.16 kbps, capable of supporting 258 QKD-secured DWDM data channels using Advanced Encryption Standard Galois/Counter Mode encryption with a key size of 256 bits (AES-256-GCM) and a key refresh rate of 1/sec. In addition, during the course of these trials, the transport of a blockchain application over this established channel was utilized as a demonstration of securing a financial transaction in transit over a quantum-secured optical channel.

It is impractical, in real-world operational environments, to require the establishment of an additional site in order to implement QKD. Rather, it is preferable that QKD devices be deployed within existing facilities. Similarly, fiber-optic links in the field are very expensive commodities; dedicating a fiber just to the quantum channel would be considered unacceptable in a real-world operational environment. Consequently, when considering a QKD solution suitable for production, the optical and quantum channels must be multiplexed on the same fiber. Even though this requirement is of the utmost importance, the deployment of high-capacity QKD-secured optical channels multiplexed with the quantum channel introduces challenges---such as polarization fluctuations as well as nonlinear effects associated with high launch powers---which have been shown to have negative impact on the performance of the quantum channel.

The experimental research described in this article was conducted purely based on real-world operational equipment, without any kind of simulation or emulation, and without the need for the deployment of additional sites dedicated to QKD. Therefore, the solution presented here is directly applicable to metro-scale operational environments. Further, in the course of this experimental research project, we conducted studies towards understanding and quantifying the impact of several degradation factors present in real-world operational environments. Specifically, we investigated the degradation of the quantum channel associated with a high-capacity QKD-secured optical channel.  Factors inducing this type of degradation include inter-channel interference (due, for example, to Raman scattering and nonlinear effects), attenuation, polarization fluctuations and distance dependency.

This study would facilitate the evolution of QKD technologies towards being suitable for implementations of quantum-secured optical channels in high-capacity mission-critical environments.

\section{Results}
\subsection{Establishing an 800 Gbps QKD-Secured Optical Channel}
\label{sec:800Gbps}
\begin{figure*}[h]
\centering
\includegraphics[width=\textwidth]{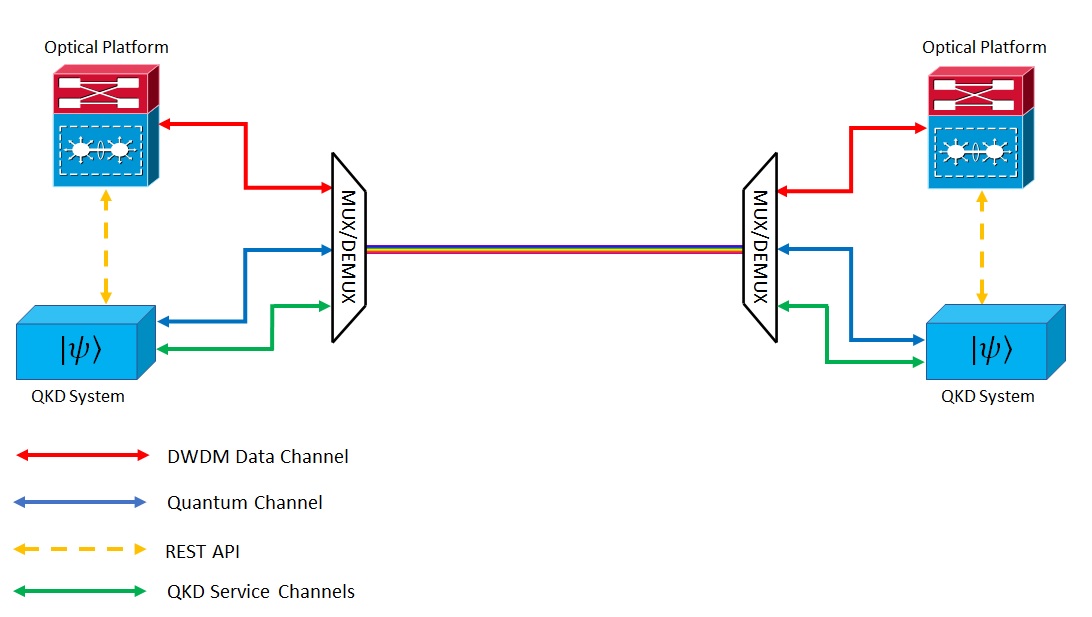}
\caption{Abstract Test-Bed Configuration: This figure shows a high-level depiction of the core test bed utilized during the experimental trials in this research study. As shown, the quantum, DWDM and service channels are all multiplexed on an optical fiber and the communication between the optical platforms and QKD systems are established using REST API.}
\label{fig:abstract-test-bed}
\end{figure*}

Figure \ref{fig:abstract-test-bed} depicts the abstract test bed configuration used in the trial. As seen from this figure, the optical data channel and the quantum channel are multiplexed on an optical fiber, while the communications between the optical and QKD devices are established using Application Programming Interfaces (APIs) on both sides. The actual realization of this test bed can be seen in Fig. \ref{fig:actual-test-bed} in the Methods section.

Upon deploying this configuration at 70 km, the quantum-secured optical channel at 800 Gbps was established, at which point the QKD systems, Alice and Bob, generated the keys. Those keys were then used to secure the optical link, which we will refer to as the Channel Under Test (CUT), using AES-256-GCM encryption over the Optical Transport Network (OTN) standard protocol, achieving 800 Gbps of quantum-secured encrypted communication on the CUT. 

\subsection{Inter-Channel Interference} \label{sec:inter-channel-interference}

After successfully establishing an 800 Gbps quantum-secured optical channel at 70 km distance, in which the channel under test (CUT) and quantum channel were multiplexed on the same fiber, the next step was to add additional optical channels. This served two purposes:

\begin{enumerate}
    \item This configuration creates a more realistic and efficient operational environment in which several optical channels are multiplexed on the C-band via DWDM.

    \item Multiplexing other optical channels results into launching higher powers into the fiber, which in turn  induces degradation impacts---such as Raman scattering and nonlinear effects---on the quantum channel.

\end{enumerate}

\begin{figure*}[h]
\centering
\includegraphics[width=140mm]{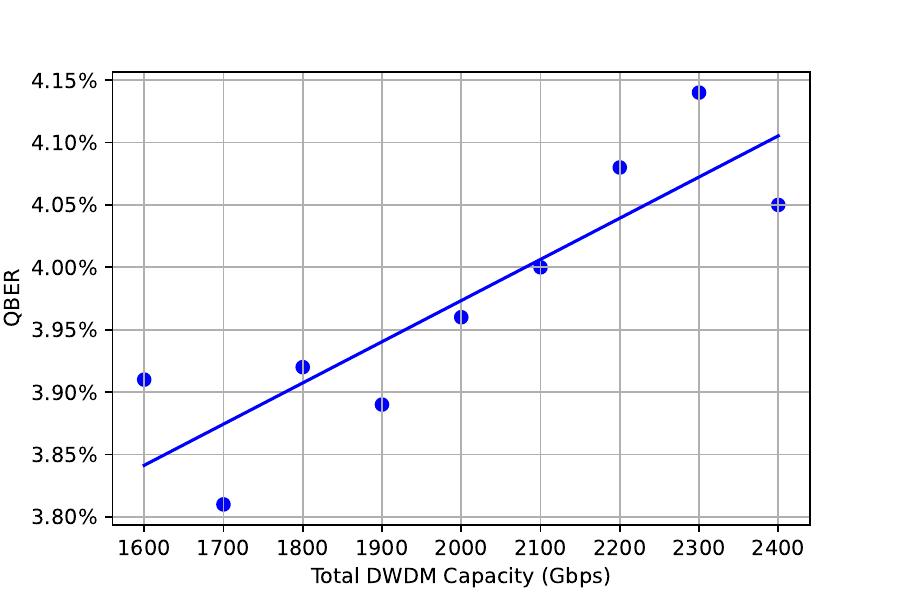}
\caption{QBER vs. Added DWDM Data: This figure shows the increase in the quantum bit error rate (QBER) as additional 100Gbps DWDM channels are added, starting with a baseline of 1600 Gbps. We choose 1600 Gbps as the baseline for this figure due to the particulars of the experimental design making it infeasible to gather sufficiently many data points between 0 and 1600 Gbps to meaningfully interpolate a trend in that region. For completeness, we include that the QBER at 800 Gbps in this trial was measured to be 3.95\%.}
\label{fig:QBER_vs_DWDM}
\end{figure*}

In total we multiplex 2.4 Tbps of DWDM data across 10 DWDM channels into the fiber link between the two endpoints. Specifically, this includes the 800 Gbps CUT, an additional 800 Gbps channel, and eight 100 Gbps channels, all distributed on the C-band. Figure \ref{fig:QBER_vs_DWDM} shows the Quantum Bit Error Rate (QBER) as a function of the total DWDM capacity. Starting at 1600 Gbps, individual 100 Gbps data channels were added to the two 800 Gbps channels in order to study the impact of inter-channel interference at high-capacities. As seen in the figure, the QBER increased linearly as a function of the total data due to the Raman scattering of the additional DWDM optical channels causing interference with the quantum channel. Specifically, in the figure we see a slight increase from 3.91\% at 1600 Gbps, to 4.04\% at 2400 Gbps, with a minimum QBER of 3.82\% observed at 1700 Gbps and a maximum QBER of 4.14\% observed at 2300 Gbps. The observed linear increase is in line with previous studies on the topic in the literature \cite{patel2014quantum,Dynes1}.  

We note that simulated results in \cite{Dynes1} expected, at a distance of 75 km, an SKR on the order of $10^3$ bps. We contrast this with our experimental results, which achieved an SKR of 66163 bps at 70 km and 30500 bps at 80 km, more than an order of magnitude increase. In \cite{Dynes1}, the simulated results were based on placing the quantum channel and approximately a total of 2.5 Tbps of DWDM channels on the C-band. In our experiments, we placed a total of 2.4 Tbps of DWDM channels generated by actual transponders on the C-band and the quantum channel on the O-band, as discussed in the Methods section of this paper. This result constitutes a logical observation due to the fact that placing the quantum channel in the O-band would create more than 200 nm band gap between the weak quantum signals and the strong DWDM channels, hence reducing the impact of Raman Scattering, although further work may be necessary to characterize the exact impact of this gap.

Further, in \cite{mao2018integrating}, the best experimental result achieved was an SKR of 5.5 kbps at 66 km, with 3.6 Tbps of total DWDM capacity over a number of 100 Gbps and 200 Gbps channels multiplexed on the C-band and with the quantum channel on the O-band. Although we note that in our case, we achieved an order of magnitude higher SKR with larger individual data-channels (800 Gbps), there were a number of differences between the experimental configurations of our studies, including total DWDM bandwidth, repetition rate, and detection methodologies. As such, it is not straightforward to compare the two results, but we include it here for completeness. 

\subsection{Distance Study}

\begin{figure*}[h]
\centering
\includegraphics[width=140mm]{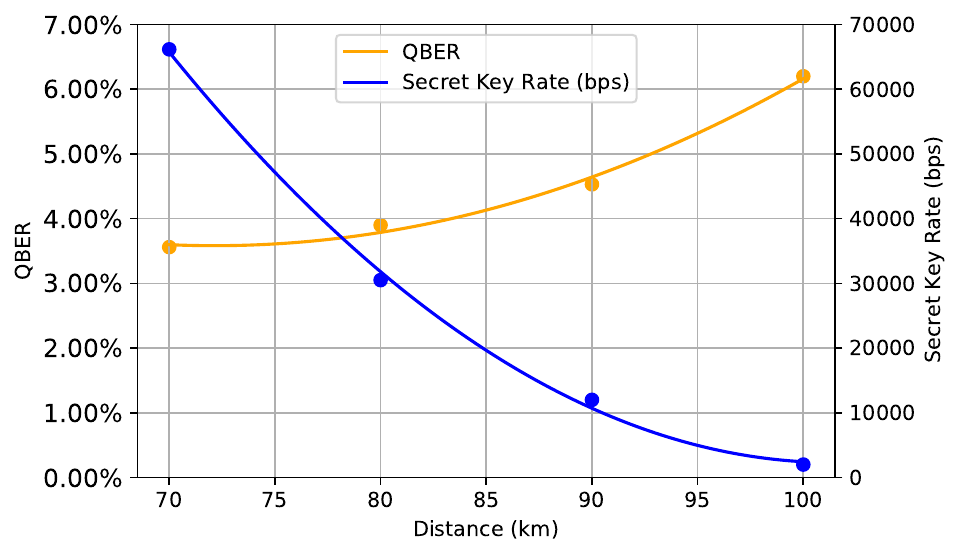}
\caption{Distance Study Results: QBER on the left axis and SKR on the right axis are plotted as the distance increases.}
\label{fig:Distance}
\end{figure*}

The focus of this part of the research study was the performance of the system as a function of the distance. Figure \ref{fig:Distance} shows the results of this study. Here, an SKR of 66.16 kbps was achieved at a 70 km distance while securing an 800 Gbps optical channel multiplexed with another 800 Gbps and eight 100 Gbps channels on the C-band. Although in this experiment there were ten optical channels present, at this rate the system has the potential to quantum-secure up to 258 optical channels using AES-256-GCM encryption while refreshing the key at a rate of 1/sec. Table \ref{tab:SKR_vs_Distance} illustrates the achieved SKR as a function of the distance. Notably, an SKR of 2 kbps was achieved at a distance of 100 km.

\begin{table}[ht]
    \centering
    \begin{tabular}{c|c}
    \hline
    \textbf{Distance (km)} & \textbf{SKR (bps)}\\
    \hline
    \hline
    70 & 66163\\
    80 & 30500\\
    90 & 12000\\
    100 & 2000\\
    \end{tabular}
    \caption{SKR vs. Distance: This table lists secret key-rates that were achieved as a result of increasing distances. }
    \label{tab:SKR_vs_Distance}
\end{table}

\subsection{Impact of Attenuation}

\begin{figure*}[h]
\centering
\includegraphics[width=140mm]{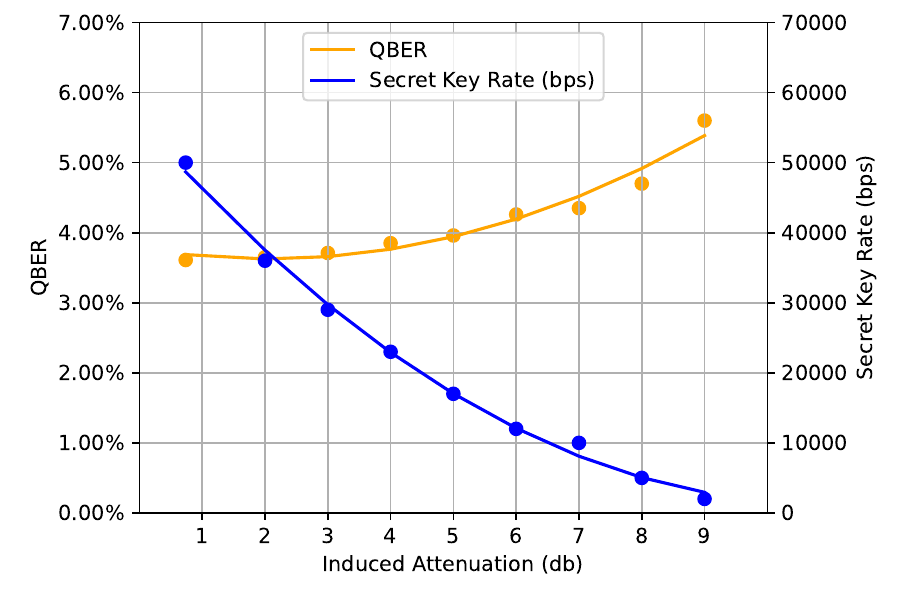}
\caption{Attenuation Study Results: Behaviour of QBER and SKR as a function of inducing additional attenuation on 70 km Fiber.}
\label{fig:VOA_Results}
\end{figure*}

Understanding the impact of increased attenuation on the QKD system in a fixed-length fiber scenario is important in order to understand the behavior of the system in a mission-critical environment. Whether the increased attenuation is as a result of a physical line issue or the presence of an eavesdropper, the network operator requires enough time to proactively investigate and avert an outage. 
In order to conduct this study, the fiber distance was set to the baseline value of 70 km and a Variable Optical Attenuator (VOA) was inserted into the quantum channel as described in the Methods section of the article. Figure \ref{fig:VOA_Results} provides the results of this study. The baseline with the VOA insertion shows a bit lower performance compared to the baseline without, i.e. a QBER of 3.61\% vs 3.56\% and an SKR of 50 kbps vs 66.16 kbps. This might be attributed to a combination of the VOA insertion loss and additional signal degradation at the interfaces introduced when connecting the signal in and out of the VOA itself.

Figure \ref{fig:VOA_Results} also shows that the maximum attenuation tolerated on top of a 70 km fiber in this configuration was 9 db, after which the key generation process stopped. This is explained by the decreasing number of photons in the quantum channel arriving at the receiver due to increasing channel loss. Once the received photon counts in the quantum channel approach the dark counts of the detector, the QBER becomes too large to distill QKD keys. This response to the induced attenuation in the form of gradual adjustment to the key generation rate---as opposed to an abrupt disruption in the process---is beneficial in mission-critical operational environment. The reason for this is that in those environments, it is vital to be aware of any degradation in the communication channel before the actual outage takes place. In that way, upon observing any degradation, the operator can proactively run a diagnosis on the system, identify the root cause (physical line issues or presence of an eavesdropper), and take necessary action before the system shuts down.

\subsection{Polarization Fluctuations}
State of Polarization (SOP) fluctuation plays a major role as a degradation factor in fiber-optic networks. This effect, mostly known as SOP Transients, is present in most operational networks as a result of vibrations in the environment such as underground fibers installed next to the railroad tracks, lightning strikes and other factors. As an example, research conducted on Optical Ground Wires (OPGW) has revealed SOP Transients on the order of 5.1 Mrad/sec as a result of lightning strikes \cite{charlton2017field}. OPGW is an aerial cable technology specifically designed for high voltage power line installations. These cables consist of a nucleus containing optical fibers and an armor, generally composed of one or more layers of aluminum wire, steel wire or aluminum-coated steel wire \cite{ITU-2009}. Therefore, any system to be deployed in operational fiber-optic networks should be able to tolerate such SOP fluctuations in order to prevent any service interruptions.

For this study, the VOA in the test bed configuration for the attenuation study was replaced with an SOP Controller. During the course of this study, it was observed that setting the SOP Controller in scrambling mode at a Stokes angular velocity of 50 rad/sec in a sustained mode caused the key generation process to stop. Although lightning strikes are usually sub-millisecond events, their impact on the quantum channel established on aerial OPGW fibers should be studied. As such, based on the findings of this trial, future research into quantifying the exact impact of extreme SOP transients on quantum channels and QKD systems are recommended.

\subsection{Routing a Blockchain Application over the Quantum-Secured Optical Channel}
 \label{sec:blockchain}

Blockchain constructions regularly make extensive use of public-key cryptography for enforcing both confidentiality and integrity through the use of encryption and digital-signature schemes, respectively, and as such have some vulnerabilities to cryptographically relevant quantum computers. In this section, we discuss the possibility of mitigating some of these weaknesses with the use of QKD-generated symmetric keys. We focus on the confidentiality for the data-in-transit of an application which we deployed over the quantum-secured channel, in an effort to provide security guarantees against a quantum-capable eavesdropper. More specifically, we build a blockchain application secure against a quantum-capable eavesdropper who wishes to compromise the security of the blockchain transactions. To this end, we installed and deployed the Liink blockchain network \cite{Liink}---a permissioned blockchain based on the Quorum protocol \cite{quorum} in use by JPMorgan Chase to process payments---and linked the nodes over quantum-secured channels. In what follows, we start by describing briefly the Liink blockchain network to give motivation for why QKD may be applicable to it, and other permissioned blockchain networks, to protect their data in transit. Following that, we also illustrate what additional tools may be deployed to enforce further security guarantees---this time against participants of the blockchain protocol rather than outside eavesdropper. Specifically, we focus on possible methods that may be used to achieve consensus and enforce transaction integrity. We choose the Liink application as a non-trivial example for the security benefits that can be obtained by integrating QKD generated keys into the security infrastructure which enables financial applications today. The experimental setup established in this study suffices to support the securing of Liink application in practical networks --- namely, where the nodes are in separate facilities connected by fiber, as in the case when the applications are hosted in a data-center-interconnects. Notably, although the improvements discussed thus far in this work may not be necessary to support a blockchain application in isolation, in practice the demonstrated ability to multiplex the quantum channel with large quantities of classical data while maintaining the necessary key rates is critical in real-world deployments of industry applications, such as the Liink application discussed.

The Liink network, as a platform utilized by financial institutions, deals with large amounts of confidential information. While the information may be intended to be read by other parties in the network, confidentiality of the data while in transit must be retained against non-participants. Currently, the confidentiality of this data in transit is protected through the use of standard public-key encryption schemes, which will not be sufficient against a quantum-capable eavesdropper.  The Liink network is an example of a permissioned blockchain network.  Therefore, there is a meaningful distinction between parties that are and are not authorized participants of the network. As such, we can differentiate those two sets of actors and their distinct security profiles. In particular, we need not ensure that transaction data remains private from participants---only from non-participants. We note that here---and everywhere else we discuss confidentiality---we only refer to the confidentiality of the data-in-transit against a non-participant, not the data-at-rest. 

The use of the quantum-secured channel allows us to achieve confidentiality against even quantum-equipped non-participants. Without the quantum-secured channel, the Liink protocol must use standard public-key cryptographic schemes to secure the confidentiality of the information in transaction payloads against eavesdroppers. Such encryption schemes rely on the standard assumptions necessary for public key cryptography, which are known to be vulnerable to attacks realizable on powerful quantum computers which are expected to become available in the future. By routing the traffic over the quantum-secured channel, we replaced the public-key encryption schemes with QKD-enabled symmetric-key cryptography ensuring that a quantum-equipped eavesdropper cannot decrypt the traffic, including sensitive transaction data. As a result, we achieve the first of our discussed guarantees: confidentiality against a non-permissioned attacker. Ensuring that a non-permissioned party cannot introduce forged messages into the network can be done through the use of symmetric-key based information-theoretic message authentication codes \cite{wegman-carter} thereby achieving our second guarantee: message integrity. As a result, we achieve both of the previously discussed guarantees: confidentiality and integrity against a \textit{non-permissioned} attacker. In addition to the increased security achieved by utilizing QKD-enabled symmetric key primitives, public key cryptography generally involves more-costly computational operations. The cost is incurred due to the particulars of the computational assumptions used in public key cryptography, and can be avoided when symmetric-key primitives are used instead. That said, we do not analyze this difference in computational costs in this work, mentioning it only for completeness here.

As the proof-of-concept blockchain network deployed in this experiment consisted of only two parties, security against non-permissioned attackers suffices to ensure the desired qualities of the blockchain network, as consensus between two parties is straightforward. In the following discussion, we sketch the modifications to the system that would be sufficient to extend the necessary security guarantees to blockchain networks composed of more than two permissioned and mistrustful participants. In any blockchain network, it is crucial that the participants are still able to achieve consensus regarding the state of the ledger, even in the presence of some number of misbehaving participants with access to less than 51\% of the total stake in the network. As a necessary step of the Liink protocol, some subset of the participants, referred to as \textit{validators}, receive, separately, transactions. For them to validate the transaction and update the ledger, they must first ensure that they all received the same transaction, reaching consensus on it. This is achieved through byzantine agreement schemes \cite{katz2006expected}, which commonly make use of broadcast protocols, which enable all of the participants to verify that they have received the same message. A natural extension of our proof of concept to more nodes encrypts the payloads via QKD-generated keys, making the construction and use of a broadcast channel costly due to the necessary cross-comparisons of a byzantine agreement protocol, but the use of optimally resilient verifiable secret-sharing schemes \cite{choudhury2021surveyVSS, beimel2011secret} can increase the efficiency, in a fully connected QKD network, and still allow the participants to reach consensus. In this approach, a transaction is split into $n$ shares according to a $k$-out-of-$n$ (optionally verifiable) secret-sharing scheme, with these $n$ shares being distributed over the QKD-secured links to the $n$ participants in the network, under the assumption that the network is fully connected. When it is their turn to validate a transaction, some $k$ participants can then form a quorum and recover the transaction from the shares. In doing so, they all necessarily agree on the same transaction, ensuring consensus \cite{share-recovery}. 

Finally, we note that it is necessary that participants of the network are confident that a transaction originates from the participant it purports to originate from. In our case, where the underlying QKD network is complete, and all parties therefore share pairwise secret keys, the participants may make use of group message-authentication codes that are information-theoretic secure to ensure integrity \cite{amiri2015unconditionally}. With a complete network we are then able to achieve consensus as well as efficiently achieve transaction integrity without the use of some underlying computational assumptions. In the case of incomplete QKD networks, the design of methods that allow validators to verify the origin of the transactions they reach consensus on without the use of such additional assumptions will be part of future work.

\section{Discussion}
To the best of our knowledge, in this research study, for the first time an 800 Gbps quantum-secured optical channel along with several other DWDM channels on the C-band and multiplexed with the QKD quantum channel on the O-band at distances up to 100 km with secure key rates relevant for use in real-world mission-critical applications has been established. At 70 km distance over Corning SMF-28 fiber, this system was able to provide a secure key rate of 66.16 kbps, which would be able to support 258 quantum-secured DWDM data channels using AES-256-GCM encryption with the key refresh rate of 1/sec. Our work, in contrast to the simulated results in \cite{Dynes1}, achieves secret key-rates an order of magnitude higher than expected at high DWDM capacities by placing the quantum channel on the O-band rather than the C-band, which is more susceptible to Raman Scattering.  As noted in \cite{Dynes1}, in the presence of high-throughput DWDM channels, the majority of the noise suffered by the quantum channel is due to Raman Scattering.  Our work suggests that one may mitigate some of that noise through a combination of (1) using the lossier but more robust O-band for the quantum channel, rather than the C-band, which is more commonly used, and (2) allocating the channels at sufficiently spaced wavelengths. While an exact characterization of the trade-offs inherent to this approach is out of scope for this work, we note that some previous analyses of this appear in the literature, for example in \cite{Eraerds2010QuantumKD}.

In addition, we demonstrated the feasibility of multiplexing the quantum channel with the 800 Gbps quantum-secure DWDM channels up to 100 km distance without any need for a third station, necessary for today's metro-scale high capacity optical networks, e.g. inter-data center interconnects. Further, during the course of these trials, the transport of a blockchain application over this established channel was utilized as a demonstration of securing a financial transaction in transit over a quantum-secured optical channel. This experimental research was conducted purely based on real-world operational equipment and no simulation or emulation was used to replace them.

Deployment of such high-capacity quantum-secured optical channels while multiplexed with the quantum channel in the real-world mission-critical operational environments would introduce challenges due to some strict requirements such as high launch powers and polarization fluctuations that would have negative impact on the performance of the quantum channel. Therefore, in the course of this research, experimental studies were conducted on the impact of several degradation factors present in the real-world operational environment on the quantum channel, including inter-channel interference (due to effects such as Raman scattering and nonlinear effects), attenuation, polarization fluctuations and distance dependency. During the course of this investigation, the entire system under test exhibited resiliency against most degradation factors. However, based on our findings, some recommendations are made towards meeting the strict requirements for deployment in mission-critical operational environments. These include future research into quantifying the exact impact of extreme SOP transients on quantum channels and QKD systems and introducing quantum channel performance degradation alarms for proactive diagnosis by the network operators to prevent outages.    

The results of this research study have addressed some of the major questions regarding the deployment of quantum-secured optical channels using QKD technology in metro-scale high-capacity mission-critical environments, such as Inter-Data Center Interconnects.

\section{Methods}
\subsection{Establishing an 800 Gbps QKD-Secured Optical Channel}

\begin{figure*}[h]
\centering
\includegraphics[width=\textwidth]{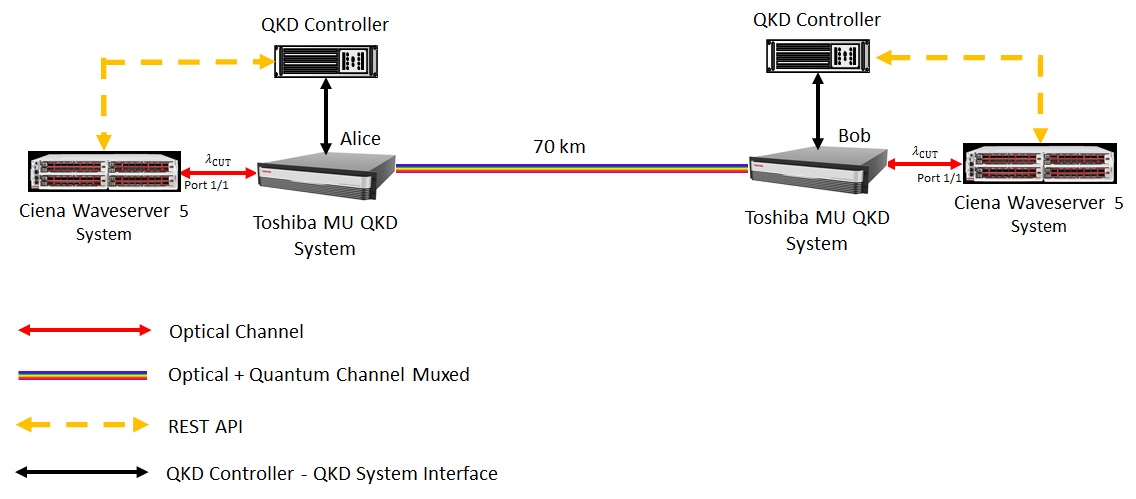}
\caption{Actual Test-Bed Configuration: This figure depicts the actual implementation of the test bed in the lab.}
\label{fig:actual-test-bed}
\end{figure*}

Figure \ref{fig:actual-test-bed} shows the actual test-bed configuration. Ciena Waveserver 5 systems (referred to as \textit{the Waveserver systems} throughout this article) were utilized to establish an 800 Gbps optical wave on the C-band as the Channel Under Test (CUT), which carries pseudo-random traffic. The Waveserver systems are able to generate and receive two 800 Gbps waves per sled and up to 4 sleds per system. However, during the course of these experimental trials, only two ports (1/1 and 1/2) were used to generate the 800 Gpbs waves as needed. For establishing the quantum channel, the Toshiba Multiplexed QKD system (referred to as \textit{the QKD system} throughout this article) was utilized. This system consists of two devices which contain transmitters, receivers and post-processing units on both sides (\textit{Alice} and \textit{Bob}), capable of establishing a quantum channel in order to generate keys based on the Decoy-State Phase-Encoded Bennett-Brassard 1984 (BB84) QKD protocol \cite{lucamarini2013efficient, bb84,yuan201810,}. These devices can also multiplex incoming optical channels on the C-band with their quantum channel generated on the O-band, and launch both on the same optical fiber. 

The 800 Gbps signal generated by the Waveserver system on the C-band on port 1/1 (the CUT) was inserted into the QKD system. The CUT was then multiplexed with the quantum channel on the O-band by the QKD system and both were transmitted over an optical fiber. On the receiver side, the reverse operation took place. Also, as shown in Figure \ref{fig:actual-test-bed}, communications between the Waveserver systems and the QKD systems were conducted via an API link between the Waveserver and the QKD Control Server device.  This communication was used by the Waveserver systems to retrieve the QKD-generated keys from a key management system on the QKD Control Server. The API link characteristics were based on the specifications of the European Telecommunications Standards Institute; ETSI GS QKD 014 \cite{ETSI}.  The two endpoints of the QKD system were connected via a 70 km Corning SMF-28 optical-fiber spool (ITU-T Recommendation G.652.D compliant).

The following list details the relevant specifications for this configuration:

\begin{itemize}
    \item $P_{\scriptsize{\textup{CUT}}}$ = 0 dBm
    \item $\lambda_{\scriptsize{\textup{CUT}}}$ = 1531.51 nm
    \item $\lambda_{\scriptsize{\textup{S1}}}$ = 1529.55 nm
    \item $\lambda_{\scriptsize{\textup{S2}}}$ = 1530.33 nm
    \item $\lambda_{\scriptsize{\textup{S3}}}$ = 1528.77 nm
    \item $\lambda_{\scriptsize{\textup{QC}}}$ = 1312.73 nm
    \item SMF-28: $\alpha_{1550}$ = 0.18 dB/km
    \item SMF-28: $\alpha_{1310}$ = 0.32 dB/km
\end{itemize}
Here, $\lambda_{\scriptsize{\textup{CUT}}}$ and $P_{\scriptsize{\textup{CUT}}}$ represent the CUT's wavelength and launch power respectively, $\lambda_{\scriptsize{\textup{QC}}}$ indicates the wavelength of the quantum channel, $\lambda_{\scriptsize{\textup{S1}}}$, $\lambda_{\scriptsize{\textup{S2}}}$, $\lambda_{\scriptsize{\textup{S3}}}$ represent the three QKD system service channels over the C-band and $\alpha$ is the attenuation coefficient.

It should be mentioned that $\lambda_{\scriptsize{\textup{CUT}}}$ was placed as close as possible to the lower edge of the C-band in order to take into account any potential interference from the QKD-secured optical channel on the quantum channel. Due to the presence of some of the required service channels on the C-band, 1531.51 nm was the closest channel we could select for this purpose. We note these required service channels are used by the systems to ensure synchronization and proper timing, as well as for the necessary post-processing procedures for QKD. In all of our discussions, these channels are multiplexed into the fiber alongside the quantum and data channels. The communication carried on the service channels is negligible when compared with the high-capacity data channels, and so, for simplicity, their inclusion is implicit in our discussions.

\subsection{Inter-Channel Interference}

As pointed out before, the Waveserver systems were able to support generation of two 800 Gbps waves per sled. In order to multiplex additional optical channels, the Ciena 6500 Photonic Line System (referred to as \textit{the photonic line system} throughout this article) and the Ciena 6500 Transponders (referred to as \textit{the transponders} hereafter) were added to the test bed. The transponders that were available in the lab were able to establish 100 Gbps waves on the International Telecommunication Union - Telecommunication (ITU-T) C-band grid. By using the photonic line system, we were able to multiplex multiple optical channels using DWDM, providing required amplification and preparing the signals for launching into the fiber. 

\begin{figure*}[h]
\centering
\includegraphics[width=\textwidth]{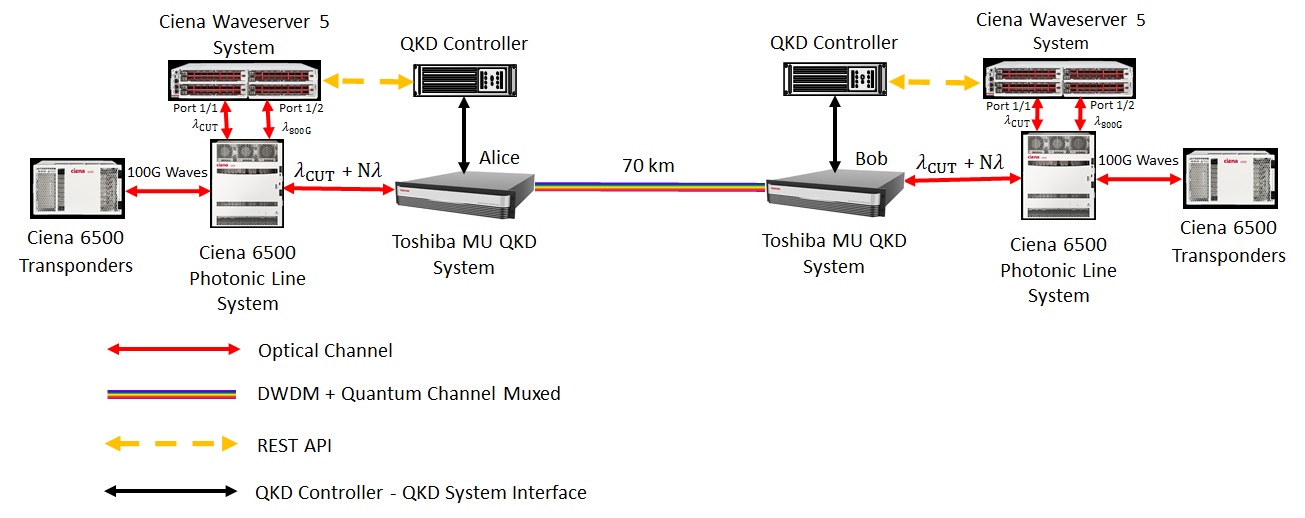}
\caption{Test-Bed Configuration for Inter-Channel Interference Study: This figure depicts the insertion of additional DWDM channels into the test bed through the Photonic Line System.}
\label{fig:All_Channels}
\end{figure*}

This configuration is depicted in Figure \ref{fig:All_Channels}. Here, the Waveserver systems generated two 800 Gbps waves, the CUT on port 1/1 and the second one on port 1/2. These waves were then inserted into the photonic line system, which multiplexed them with eight 100 Gbps DWDM channels generated by the transponders on the C-band. After passing through an amplifier module on the photonic line system, these multiplexed DWDM channels (the two 800 Gbps and eight 100 Gbps channels) were inserted into the QKD system, which in turn multiplexed them with the quantum channel running on the O-band. As before, the two endpoints of the QKD system were connected via 70 km of Corning SMF-28 fiber. It should be noted that, in Figure \ref{fig:All_Channels}, $N\lambda$ represents all other wavelengths that $\lambda_{\scriptsize{\textup{CUT}}}$ has been multiplexed with on the C-band. 

As before, these optical channels were also carefully placed as close as possible to the lower edge of the C-band in order to maximize the potential of any inter-channel interference on the quantum channel. Table \ref{tab:Channel_Lineup} provides the channel lineup for this configuration.

Since this test-bed configuration is the closest to the actual operational environments in the field, the rest of the experiments were conducted over this configuration.

\begin{table}[ht]
    \centering
    \begin{tabular}{l|c}
    \hline
    \textbf{Channel} & \textbf{Wavelength (nm)}\\
    \hline
    \hline
    CUT (800 Gbps) & 1531.51\\
    Second 800 Gbps & 1532.68\\
    100 Gbps No. 1 & 1533.86\\
    100 Gbps No. 2 & 1534.25\\
    100 Gbps No. 3 & 1534.64\\
    100 Gbps No. 4 & 1535.04\\
    100 Gbps No. 5 & 1535.43\\
    100 Gbps No. 6 & 1535.82\\
    100 Gbps No. 7 & 1536.22\\
    100 Gbps No. 8 & 1536.61\\
    \end{tabular}
    \caption{Channel Lineup: This table provides the wavelengths of all the DWDM channels inserted into the test bed for the Inter-Channel Interference study.}
    \label{tab:Channel_Lineup}
\end{table}

\subsection{Distance Study}
In this part of the study, different fiber spools were used in the test bed specified in Figure \ref{fig:All_Channels} to analyze the impact of increasing the distance on the system performance. All fibers were of type Corning SMF-28 G.652.D. It should also be mentioned that for 80 km to 100 km distances, the launch power on the Waveserver systems were increased to 2 dBm in order to keep the 800 Gbps optical signals within the acceptable power levels at the receivers.

\subsection{Impact of Attenuation}

\begin{figure*}[h]
\centering
\includegraphics[width=\textwidth]{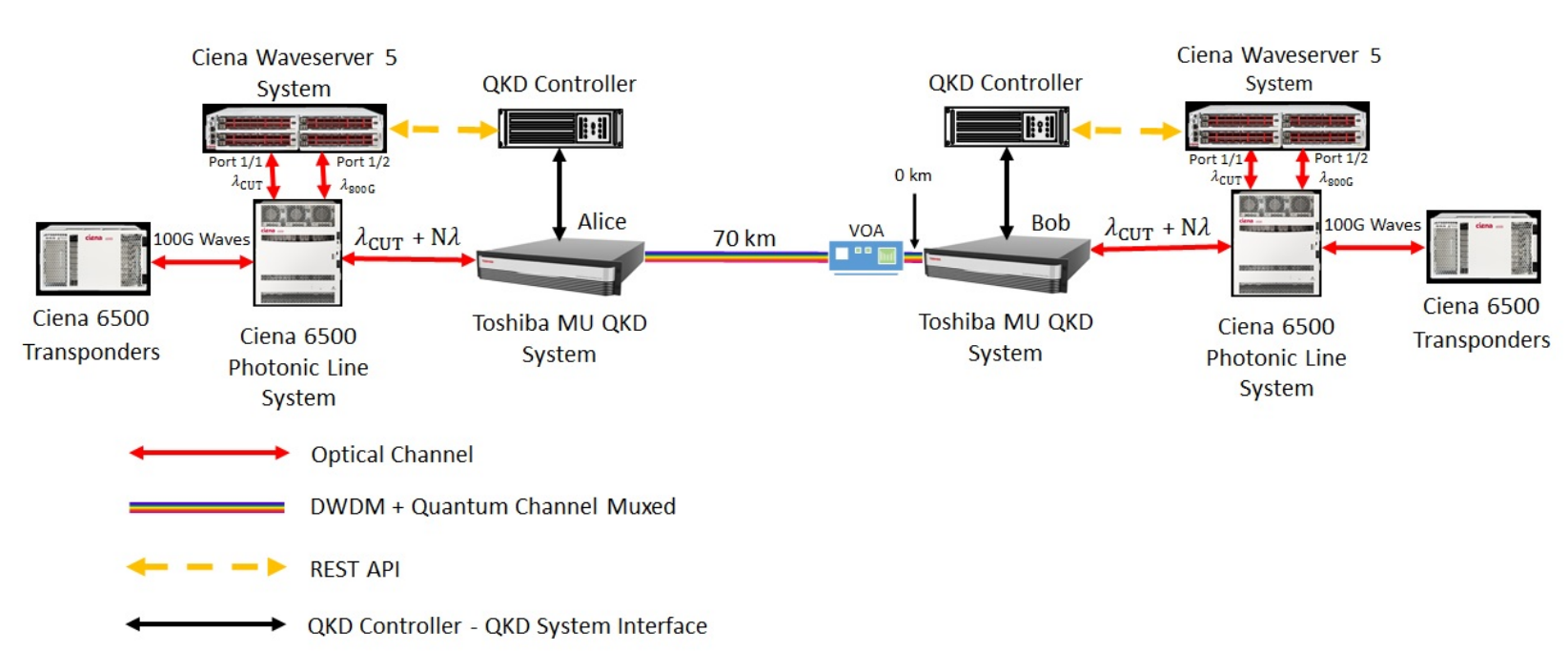}
\caption{Attenuation Study Configuration: Test bed configuration for the attenuation study depicting insertion of the VOA unit into the fiber optic link.}
\label{fig:VOA}
\end{figure*}

As shown in Figure \ref{fig:VOA}, the VOA insertion point was right between the termination point of the 70 km fiber and the receiver port on Bob's QKD system. The VOA had an insertion loss of 0.737 dB and the target wavelength was set to 1310 nm in order to directly impact the quantum channel. The VOA attenuation value was increased at 1 dB increments on an hourly basis and measurements were taken.

\subsection{Routing a Blockchain Application over the Quantum-Secured Optical Channel}
\begin{figure*}[h]
\centering
\includegraphics[width=\textwidth]{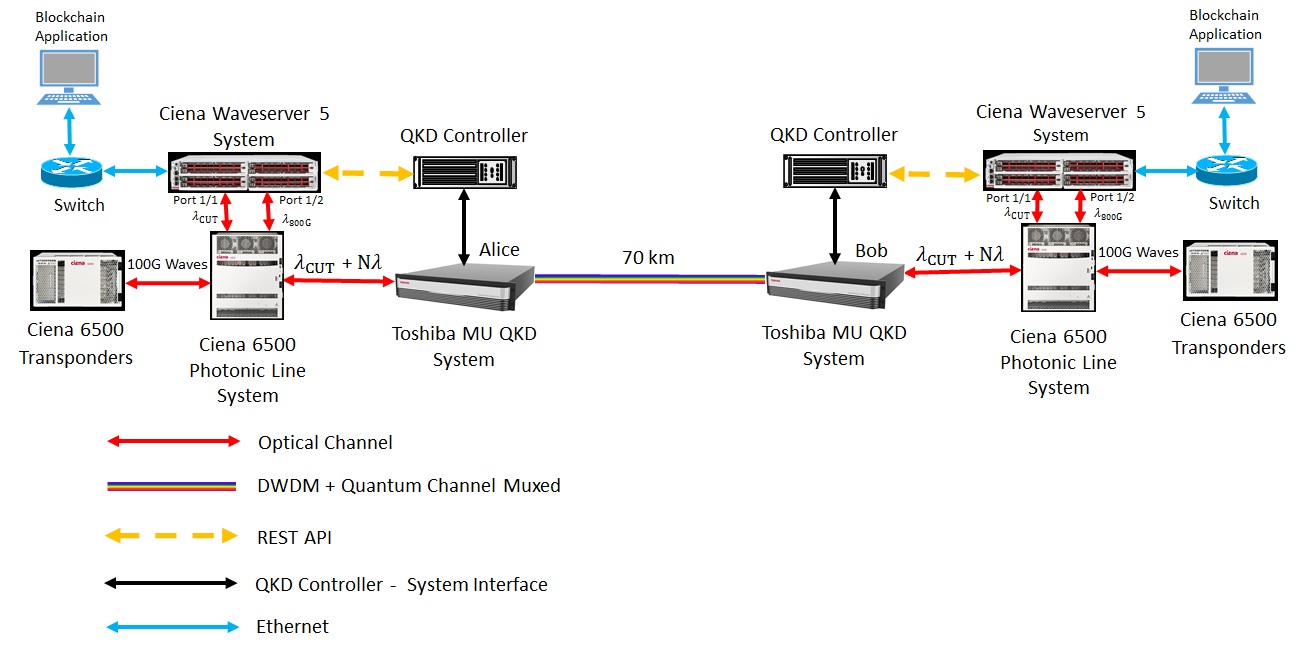}
\caption{Blockchain Application Transport Test-Bed Configuration: This figure shows the addition of the blockchain application servers and switches to the test bed for the blockchain transport experiment.}
\label{fig:blockchain}
\end{figure*}

In the experimental trial, as shown in the lab test bed depicted in Figure \ref{fig:blockchain}, the Liink application and transaction data were transmitted between two virtual machines (VMs) across the 800 Gbps quantum-secured optical channel (the CUT). The VMs were connected to switches that were in turn connected to the Waveserver systems via 100 Gbps Ethernet links, the output of which, carrying the traffic, was then sent over the CUT. A lightweight version of the Liink application was setup for this experiment. All communications between the two blockchain nodes Alice and Bob, including Peer Discovery, and transaction propagation, were conducted over the quantum-secured optical channel. Upon initiating a transaction by node Alice, a listener process on node Bob confirmed the receipt of the transaction along with its addition into the block. Once included in the block, the transaction was considered confirmed by the two involved parties. During this experiment, the application did not encrypt the JSON payload in the transaction that it published on blockchain since protection was provided by the quantum-secured channel.

\section*{Disclaimer}

This paper was prepared for information purposes by the teams of researchers from the various institutions identified above, including the Future Lab for Applied Research and Engineering (FLARE) group of JPMorgan Chase Bank, N.A..  This paper is not a product of the Research Department of JPMorgan Chase \& Co. or its affiliates.  Neither JPMorgan Chase \& Co. nor any of its affiliates make any explicit or implied representation or warranty and none of them accept any liability in connection with this paper, including, but limited to, the completeness, accuracy, reliability of information contained herein and the potential legal, compliance, tax or accounting effects thereof.  This document is not intended as investment research or investment advice, or a recommendation, offer or solicitation for the purchase or sale of any security, financial instrument, financial product or service, or to be used in any way for evaluating the merits of participating in any transaction.

\section*{Acknowledgements}
The authors would like to thank the following people for their invaluable contributions and support to this project: Lori Beer, Andrew J.~Lang, Charles Lim, George A.~Sherman, Umar Farooq, Matthew Tice, Beata Shahriari, Ruslan Shaydulin, Myron Wieneke, Stephen Gannon, Charles Carley, Douglas B.~David, Michael P.~Lombardi, Mike Rutigliano, Paula Varas, Stephen I.~Winer, Jennifer H.~Lavoie, Daniel Cantor, Hargun Kalsi, Michael Hattery, Alejandro Poggi, Chaddy Huussin and Emily Mullins from JPMorgan Chase; Terry Cronin from Toshiba; and Patrick Scully, Daniel McCarthy, Chris Howard, John P.~Ferrel, Sebastien Gareau and Yves Cayer from Ciena.

\printbibliography
\end{document}